\documentclass[conference]{IEEEtran}
\usepackage{cite}
\usepackage{amsmath}
\usepackage{amsmath,amssymb,amsfonts}
\usepackage{pifont}
\usepackage{algorithm}
\usepackage{algorithmic}
\usepackage{graphicx}
\usepackage{textcomp}
\usepackage{xcolor}
\usepackage[hidelinks]{hyperref}
\usepackage{listings}
\usepackage{enumitem}
\usepackage{balance}
\usepackage[T1]{fontenc}
\usepackage[group-separator={,},group-minimum-digits=4]{siunitx}
\usepackage{booktabs}
\usepackage{multirow}
\usepackage{tabularx}
\usepackage{subcaption}
\definecolor{codegreen}{rgb}{0,0.6,0}
\definecolor{codegray}{rgb}{0.5,0.5,0.5}
\definecolor{codepurple}{rgb}{0.58,0,0.82}
\definecolor{backcolour}{rgb}{0.95,0.95,0.92}

\ifCLASSINFOpdf
\else
\fi

\makeatletter
  \def\sectionautorefname{\S\@gobble}
  \def\subsectionautorefname{\S\@gobble}
  \def\subsubsectionautorefname{\S\@gobble}

  \newcommand{\removelatexerror}{\let\@latex@error\@gobble}
\makeatother

\newif\iffinal
\finaltrue
\iffinal
    \newcommand{\sicheng}[1]{}
    \newcommand{\zhuozhao}[1]{}
    \newcommand{\kyle}[1]{}
    \newcommand{\yifei}[1]{}
    \newcommand{\ryan}[1]{}
    \newcommand{\haochen}[1]{}
    \newcommand{\ian}[1]{}
    
    \newcommand{\change}[1]{{\textcolor{black}{#1}}}

\else
    \newcommand{\sicheng}[1]{{\textcolor{purple}{ Sicheng: #1 }}}
    \newcommand{\zhuozhao}[1]{{\textcolor{blue}{ ZZ: #1 }}}
    \newcommand{\kyle}[1]{{\textcolor{red}{ Kyle: #1 }}}
    
    \newcommand{\yifei}[1]{{\textcolor{magenta}{ Yifei: #1 }}}
    
    \newcommand{\ryan}[1]{{\textcolor{magenta}{ Ryan: #1 }}}
    \newcommand{\haochen}[1]{{\textcolor{brown}{ Haochen: #1 }}}
    \newcommand{\ian}[1]{{\textcolor{blue}{ Ian: #1 }}}
    
    \newcommand{\change}[1]{{\textcolor{blue}{#1}}}
      
\fi

\newcommand{\name}{\textsc{Wrath}}

\begin{document}
\title{\name{}: Workload Resilience Across Task Hierarchies in Task-based Parallel Programming Frameworks }

% \author{
% \IEEEauthorblockN{Anonymous Author(s)}
% \IEEEauthorblockA{}
% }

\author{
    \IEEEauthorblockN{
        Sicheng Zhou\IEEEauthorrefmark{1},
        Zhuozhao Li\IEEEauthorrefmark{1},
        Valérie Hayot-Sasson\IEEEauthorrefmark{2}, 
        Haochen Pan\IEEEauthorrefmark{2}, 
        Maxime Gonthier\IEEEauthorrefmark{2}, \\
        J. Gregory Pauloski\IEEEauthorrefmark{2},
        Ryan Chard\IEEEauthorrefmark{3}, 
        Kyle Chard\IEEEauthorrefmark{2}\IEEEauthorrefmark{3}, 
        Ian Foster\IEEEauthorrefmark{3}\IEEEauthorrefmark{2}
    }
    \IEEEauthorblockA{
        \IEEEauthorrefmark{1}Department of Computer Science, Southern University of Science and Technology, Guangdong, China\\
    }
    \IEEEauthorblockA{
        \IEEEauthorrefmark{2}Department of Computer Science, University of Chicago, Chicago, IL, USA\\
    }
    \IEEEauthorblockA{
        \IEEEauthorrefmark{3}Data Science and Learning Division, Argonne National Laboratory, Lemont, IL, USA\\
    }
}

\maketitle

\begin{abstract}
Failures in Task-based Parallel Programming (TBPP) can severely degrade performance and result in incomplete or incorrect outcomes. 
    Existing failure-handling approaches, including reactive, proactive, and resilient methods such as retry and checkpointing mechanisms, often apply uniform retry mechanisms regardless of the root cause of failures, failing to account for the unique characteristics of TBPP frameworks such as heterogeneous resource availability and task-level failures. 
    To address these limitations, we propose \name{}, a novel systematic approach that categorizes failures based on the unique layered structure of TBPP frameworks and defines specific responses to address failures at different layers.
    \name{} combines a distributed monitoring system and a resilient module to collaboratively address different types of failures in real time.   
    The monitoring system captures execution and resource information, reports failures, and profiles tasks across different layers of TBPP frameworks. The resilient module then categorizes failures and responds with appropriate actions, such as hierarchically retrying failed tasks on suitable resources.
    %reschedules failed tasks hierarchically based on the gathered data. 
    % Our contributions include a survey of failure types in distributed task-based applications, techniques for detecting failures across different stack layers, and methods to enhance makespan and system efficiency.
    Evaluations demonstrate that \name{} significantly improves TBPP robustness, tripling the task success rate and maintaining an application success rate of over 90\% for resolvable failures. Additionally, \name{} can reduce the time to failure by 20\%-50\%, allowing tasks that are destined to fail to be identified and fail more quickly.
\end{abstract}


% \fi
\IEEEpeerreviewmaketitle

\thispagestyle{plain}
\pagestyle{plain}

\section{Introduction} \label{sec:intro}

% \valerie{I would really recommend starting the first paragraph on failures in task execution and then present the scope at which we will address failures.}
% \sicheng{We should define and describe TBPP before mentioning failures in TBPP, shouldn't we?}\valerie{The issue is you want people to have a general sense of what your paper is about in the first paragraph. If you start off with a lot of description, readers will lose interest or not understand the goal of the paper properly. I would start with your second paragraph, and leave TBPP descriptions in background}\sicheng{Done, please check.}

% Task-Based Parallel Programming (TBPP) is a paradigm that emphasizes the execution of independent tasks concurrently across multiple processing units. Each task is a self-contained unit of work with well-defined inputs and outputs, allowing for dynamic scheduling and efficient resource utilization. This concept has become pivotal in modern applications, enabling the efficient execution of concurrent operations by decomposing large computations into smaller, manageable tasks. The TBPP frameworks, such as Dask~\cite{rocklin2015dask}, Parsl~\cite{babuji2019parsl}, and Ray~\cite{moritz2018ray}, optimize resource utilization, improve performance, and simplify the development process by abstracting low-level threading details and offering high-level constructs for parallelism, providing developers with robust tools to harness the full potential of heterogeneous computing resources. 

Task-based parallel programming (TBPP) is a programming paradigm in which a computational workload is divided into discrete units of work called tasks. 
These tasks can execute concurrently on the same or different computing nodes, subject to constraints resulting from shared data and communication.
Particularly as parallel systems and applications increase in complexity and scale, it becomes crucial to be able to detect and recover from various forms of task failure, such as can result from hardware malfunctions, software bugs, and incompatible environments (e.g., due to different libraries, system modules, and even Python versions). %Task execution failures represent a critical challenge in Task-based Parallel Programming (TBPP), where the concurrent execution of independent tasks is subject to various types of failures, including hardware malfunctions, software bugs, and unexpected environmental conditions. Such failures can disrupt the overall computation process, resulting in incomplete or incorrect results and potentially leading to significant performance degradation. As the complexity and scale of parallel systems increase, the need for effective mechanisms to detect and recover from such failures becomes paramount to ensure reliability and efficiency.

Commonly used TBPP frameworks, such as Dask~\cite{rocklin2015dask}, Parsl~\cite{babuji2019parsl}, and Ray~\cite{moritz2018ray}, incorporate various basic resilience mechanisms to mitigate task failures, such as \emph{task retry}, in which the system automatically attempts to rerun a failed task, often with a configurable number of retries, and \emph{checkpointing}, which involves saving the state of a computation periodically and resuming from the last checkpoint.
%rather than starting from scratch in case of failure.
However, these mechanisms are often insufficient for handling the complexities inherent in large-scale distributed systems, where failures may stem from heterogeneous resource availability or task-specific issues. Hence, failures in TBPP applications are diverse, distributed, and may occur at multiple levels---from application-specific bugs, to resource starvation or failures in the distributed hardware on which tasks are executed. 
Basic resilience methods typically apply uniform retry mechanisms without distinguishing between failure types.
This one-size-fits-all approach can lead to inefficient responses, such as needlessly retrying tasks with application-specific bugs that will fail again or overlooking systemic issues that require a more coordinated recovery effort.

In this paper, we explore methodologies and techniques for handling failures in TBPP frameworks.  We review failures at different levels of TBPP frameworks, define categories of failures with similar characteristics and necessary responses, examine existing approaches to failure management, and propose improvements and best practices to enhance the robustness of TBPP frameworks. 
We present a new approach called \name{} (\underline{W}orkload \underline{R}esilience \underline{A}cross \underline{T}ask \underline{H}ierarchies)
for addressing failures in TBPP applications. 
% We focus specifically on the unique types of failures that can occur at different levels across a distributed ecosystem. 
\name{} categorizes failures at four layers of TBPP frameworks and defines specific responses to address failures at different levels. 
%, and proposes to categorize failures based on the TBPP layers.
\name{} includes a hierarchical \textit{monitoring system} and an intelligent \textit{resilience module}. 
Specifically, the monitoring system leverages distributed monitoring agents to gather valuable information across the hierarchies of TBPP frameworks.
Meanwhile, the resilience module employs new failure categorization methods to identify failures from monitoring information and map them to appropriate handling mechanisms, such as immediate failure responses or retries.
Additionally, the resilience module implements a \emph{hierarchical retry} mechanism that dynamically retries failed tasks across different resource pools, thereby increasing the likelihood of successful task execution.

The novelty of \name{} lies in its introduction of a framework that categorizes failures based on four distinct layers of TBPP frameworks, contrasting with existing approaches that typically apply a uniform retry mechanism for all types of failures. Additionally, \name{} considers the hierarchical nature of TBPP frameworks by proposing a distributed monitoring system and a hierarchical retry mechanism, enabling more effective and tailored responses to diverse failure scenarios.

The key contributions of this paper include:
\begin{itemize}
    \item A comprehensive survey of failure types in TBPP frameworks and categorization of these types  at different layers of the stack;
    \item Implementation of a distributed monitoring system for real-time data collection across the TBPP stack, facilitating a more informed and adaptive response to failures;
    \item Development of a resilience module based on the proposed failure categorization methods, which maps identified failures to appropriate handling mechanisms, as well as a hierarchical retry mechanism that dynamically reallocates failed tasks to different resource pools;
    \item A thorough evaluation of \name{} using a benchmark system with real TBPP applications to demonstrate the effectiveness of \name{} in terms of success rate, overhead, and ``fail fast'' for non-resolvable failures.
\end{itemize}

The rest of this paper is as follows: \S\ref{sec:bg} introduces background and motivation; \S\ref{sec:characterize} categorizes failures across the layers of TBPP frameworks; \S\ref{sec:detect} presents methods to detect failures; \S\ref{sec:respond} describes ways to respond to failures; \S\ref{sec:implementation} provides the detailed implementation of \name{}; \S\ref{sec:evaluation} presents experiments to evaluate our solution; \S\ref{sec:related_work} discusses related work; and \S\ref{sec:conclusion} concludes the paper with future remarks.

\section{Background and Motivation} \label{sec:bg}

In this section, we introduce the fundamentals of task-based parallel programming (TBPP) frameworks, define failures, and highlight the motivations for our approach.
We assume here, as is common~\cite{dagman,di2017nextflow,rocklin2015dask,airflow, babuji2019parsl}, atomic tasks: i.e., tasks that either complete and generate output or fail with no output.
%\textit{idempotent} meaning that a task can be executed more than once without problems.
This property means that resilience can be achieved by re-executing failed tasks.

\subsection{TBPP Frameworks}
TBPP is a programming paradigm that divides large computational problems into smaller units known as \emph{tasks}. TBPP frameworks facilitate the definition of tasks with explicit dependencies and enable their scheduling across various computing resources. The interleaved execution of these tasks may be subject to constraints arising from control- and data-flow dependencies~\cite{thoman2018taxonomy}.

Frameworks such as Dask, Parsl, and Ray abstract low-level parallelization details and offer high-level constructs that enable developers to express parallelism without needing to manage the underlying hardware directly. These frameworks typically include a task scheduler that dynamically assigns tasks to available computing resources based on task priority, dependencies, and resource availability.

\begin{figure}[h]
    \centering
    \includegraphics[page=1, width=\linewidth, clip=true, viewport=0 220 460 400]{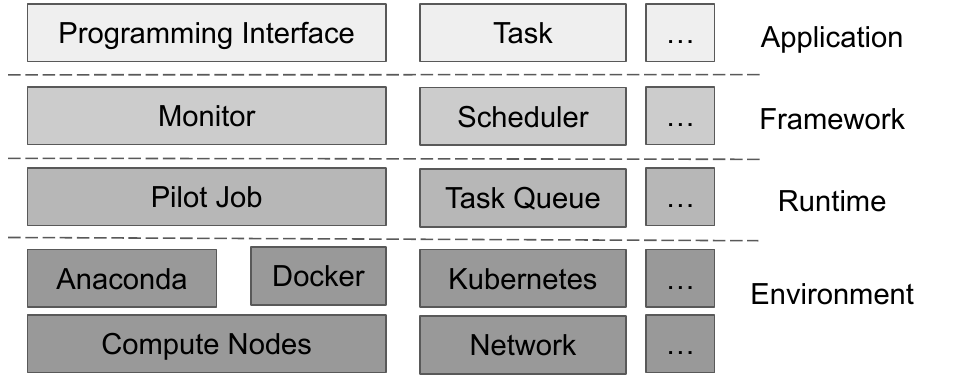}
    \caption{Typical architecture of TBPP frameworks. In the \textit{Application Layer}, users define applications into tasks using the provided programming interfaces. The \textit{Framework Layer} orchestrates the execution of tasks. The \textit{Runtime Layer} allocates resources to tasks. The \textit{Environment Layer} manages the underlying infrastructure and package dependencies. 
    % \kyle{Note to update figure to Framework layer}
    }
    \label{fig:arch}
     \vspace{0em}
\end{figure}

While TBPP frameworks offer numerous advantages, such as simple programming models, high performance, and scalability, detecting and responding to failures is difficult as tasks are executed on heterogeneous distributed computing resources.
Failures can occur at four layers (i.e., application, framework, runtime, and environment layers, as shown in \autoref{fig:arch}) of a TBPP system~\cite{thoman2018taxonomy}, each presenting challenges for fault detection, management, and recovery.

The \textbf{Application Layer} is where tasks are defined. 
This layer involves the coding of tasks, including the algorithms, data structures, and logic that will be executed. Tasks at this layer %are typically designed to be as independent as possible, but 
may have explicit dependencies on other tasks.
While different TBPP frameworks
% (e.g., Pegasus, Dask, Ray, RADICAL-Pilot, Parsl) 
enable task definition and construction using different interfaces (e.g., YAML, Java, Python), they share common features such as defining task dependencies as directed acyclic graphs (DAGs).  % handle failure with built-in retry mechanisms.

% The \textbf{Application Layer} refers to the topmost layer where the user's code or application logic resides. The system provides API to users, allowing them to specify the workload that will be executed in parallel, including tasks, input, output, and the dependency between the executables. In Pegasus, programmatic APIs in Python, Java, and Perl are provided for users to define the input/output datasets and executables as logical identifiers \cite{Pegasus}. For Parsl, Python functions are annotated as Apps with datasets and dependency specifications \cite{babuji2019parsl}. Ray lets users specify remotely executed tasks and retrieve results by using ray.get() \cite{moritz2018ray}.

The \textbf{Framework Layer} orchestrates the execution of tasks defined in the application layer.
Most TBPP frameworks rely on a \emph{central manager} to manage the task dependencies, scheduling, monitoring, and failure handling, although (even fully) decentralized approaches can be used~\cite{iamnitchi2000problem}.
This layer ensures that tasks are executed in the correct sequence and that computational resources are utilized efficiently. The framework layer may respond to failures from the runtime layer by retrying task execution. % For example, in Parsl, the framework offers a configurable retry handler that can be configured to retry tasks a predefined number of times.  %using different scheduling mechanisms. 

%Dask implemented last-in, first-out schedulers to ensure a small memory footprint \cite{rocklin2015dask}. Ray proposed the global control store which maintains the entire control state of the system, the two-level hierarchical scheduler, and the distributed object store which stores the inputs and outputs of every task \cite{moritz2018ray}. 
% \sicheng{Dask, Ray, Workqueue}

The \textbf{Runtime Layer} is responsible for managing task execution on underlying computational resources. TBPP frameworks often rely on the \emph{pilot job model}~\cite{pilot}, in which placeholder jobs are submitted to computing resources to initialize the execution environment and hold the resources. Usually, the pilot job will start a \emph{node manager} process on each node in the job, which is responsible for receiving tasks and assigning them to \emph{worker} processes responsible for executing the tasks.
% Once the pilot job is running, the pilots pull tasks (often referred to as payloads) from a task queue and execute them. 

% The glideinWMS automatically submits pilot jobs to the Grid as long as there are jobs that could potentially run in the queues, trying to maximize the number of completed jobs while minimizing the wasted resources \cite{glideinWMS}. DIRAC uses central TaskQueues, sorted groups of payloads waiting for execution with identical requirements to the possible Pilot Job requesting a match, to provide transparent and efficient access to the underlying computing resources, allowing an overall optimization of the resource usage for the complete user community \cite{DIRAC}.

The \textbf{Environment Layer} includes the underlying infrastructure and the runtime environment in which the application, framework, and runtime system operate.
Modern TBPP frameworks often leverage containers (e.g., Docker or Singularity) or environment management software (e.g., conda or virtual environment) to allow developers to create consistent and portable environments. 
%that avoid software conflicts and allow applications to perform reliably regardless of where they are executed. 
This layer also enhances the reproducibility of computational tasks, a key requirement in scientific computing and other fields where results must be verified and validated across different platforms.

\begin{table*}[ht]
\caption{Failure types and detection strategies. FTL stands for Failure Taxonomy Library; RP for Resource Profiling; and RC for depending on the Root Cause.}
\centering
\begin{tabularx}{\textwidth}{lllcc}
\toprule
\multicolumn{1}{l}{\textbf{\begin{tabular}[c]{@{}l@{}}Layer \&\\ Failure Type\end{tabular}}} & \multicolumn{1}{l}{\textbf{Example Root Cause}} & \multicolumn{1}{l}{\textbf{Examples \& Description}} & \multicolumn{1}{l}{\textbf{\begin{tabular}[c]{@{}l@{}}Detection\\ Strategy\end{tabular}}} & \multicolumn{1}{l}{\textbf{\begin{tabular}[c]{@{}l@{}}Is Failure\\ Retriable\end{tabular}}} \\ \hline
\multirow{3}{*}{\begin{tabular}[c]{@{}l@{}}Application Layer\\ (User Failures)\end{tabular}} 
 & Syntax Errors & Mistakes that violate programming language syntax. & FTL & No \\
 & Logic Errors & Array index out-of-bounds or incorrect data types. & FTL & No \\
 & Random Seed Errors & Molecule Design initialization issue. & / & Yes \\ \hline
\multirow{3}{*}{\begin{tabular}[c]{@{}l@{}}Framework Layer\\ (System Failures)\end{tabular}} & Monitor Loss & Task overseeing component becomes unavailable. & FTL & Yes \\
 & Manager Loss & The component responsible for managing tasks fails. & FTL & Yes \\ 
 & Dependency Failures & Single task failure causes multiple dependent tasks to fail. & RC & RC \\\hline
\multirow{2}{*}{\begin{tabular}[c]{@{}l@{}}Runtime Layer\\ (Resource Failures)\end{tabular}} & Resource Starvation & Insufficient memory or CPU to execute a task. & RP & Yes \\
 & Pilot Job Init. Failures & The pilot job fails to initialize correctly. & RP & Yes \\ \hline
\multirow{2}{*}{\begin{tabular}[c]{@{}l@{}}Environment Layer\\ (Hw. \& Env. Failures)\end{tabular}} & Hardware Shutdown & Components such as servers or network devices fail. & FTL + RP & Yes \\
 & Runtime Env. Mismatches & Missing required software versions or libraries in env. & FTL & No \\ \bottomrule
\end{tabularx}
\label{tab:failures}
\end{table*}

\subsection{Motivation and the \name{} Approach}

As mentioned in \S\ref{sec:intro}, existing TBPP frameworks typically rely on \emph{retry} and \emph{checkpointing} approaches to ensure resilience. However, these approaches often adopt a \emph{flat} structure, meaning they treat all failures uniformly without considering their context or root causes within the TBPP layers. This limitation prevents the identification and resolution of failures in a manner that is tailored to the specific characteristics and complexities of the system.
Therefore, we are motivated to design a failure-handling system that considers the unique characteristics of TBPP frameworks, including the heterogeneity and layered structure.

In this paper, we introduce \name{}, a failure-handling approach that aims to enable resilient computing for TBPP frameworks. 
By resilient computing, we refer to the ability of a computing system to continue functioning properly in the presence of failures. 
%In this paper, we introduce our approach, \name{}. 
\name{} proposes to categorize various failures based on the layers of TBPP frameworks. 
To monitor the failures in different layers, \name{} integrates a hierarchical monitoring system to gather execution and resource data and report failures as they occur. 
Based on the categorization methods and the monitoring system, \name{} designs a dynamic resilience module that can intelligently retry failed tasks on the most appropriate resource pools.

In the next three sections, we will detail how \name{} characterizes failures in TBPP frameworks, the design of the hierarchical monitoring system, and the implementation of the dynamic resilience module.

% we introduce \name{}, a novel framework that integrates a distributed monitoring system with a resilient module, aiming to address various failure types in TBPP in real time. 
% The monitoring system continuously gathers execution and resource data, profiling tasks and reporting failures as they occur. 
% The resilient module classifies failure types and applies tailored responses, such as hierarchical task rescheduling based on real-time insights. 
% This cooperative approach between monitoring and resilience mechanisms ensures timely and accurate failure detection, enabling TBPP systems to recover efficiently and sustain high performance even under adverse conditions.

\section{Characterization of Failures in TBPP} \label{sec:characterize}

Despite the robustness of modern TBPP frameworks, failures can occur at multiple levels, disrupting execution and potentially leading to incorrect results, wasted resources, and performance degradation. Understanding the nature of failures is crucial for designing resilient systems that can either recover from them or minimize their impact effectively. 

% \maxime{The subsection below was hard to grasp in my opinion.
% I propose to write the following instead: 
% "
% We define a correct service as the submitted function being successfully delivered.
% A \textbf{failure} is defined as a deviation between the delivered function and the submitted function.
% Failures can be explicit, such as when a program crashes, but can also be silent, such as when results are incorrect.
% The deviation from the correct service is called an \textbf{error}.
% The hypothesized cause of an error is denoted by a \textbf{fault}. \cite{concepts}.
% "}
% \sicheng{Most of the words in this paragraph are cited from \cite{concepts} so it's not "We define". What about this, I only modified the 1st sentence:}
% \maxime{Yes looks good to me now :)}
% A correct service is the successfully delivered submitted function. A \textbf{failure} is defined as a deviation between the delivered function and the submitted function. Failures can be explicit, such as when a program crashes, but can also be silent, such as when results are incorrect. The deviation from the correct service is called an \textbf{error}. The hypothesized cause of an error is denoted by a \textbf{fault}\cite{concepts}.

\subsection{Failure Root Causes} \label{subsection:root_cause}
% \haochen{Failures in different layers, definition, examples, impact}

Different types of failures may occur at different layers in TBPP frameworks. We summarize them in \autoref{tab:failures}.

Failures at the \textbf{Application Layer} are 
\textbf{User Failures} due to mistakes or incorrect assumptions made by users when writing their application code and tasks. 
These failures may be incorrect results, crashes, or inefficient execution. 
Typical causes include \textit{Syntax Errors}, where mistakes in the code violate programming language syntax rules; \textit{Logic Errors}, such as incorrect use of loops or mathematical calculations, accessing out-of-bounds array indices, or incorrect data types; and \textit{Random Seed Errors}, where the failure is sporadic. An example of Random Seed Error can be seen in a molecule design application~\cite{MolDesign}, during the first period of simulation in which molecule assumptions are generated randomly for further calculation. These assumptions can cause errors in simulation and subsequent processing. But after regeneration, the errors may resolve.

Failures at the \textbf{Framework Layer} are TBPP \textbf{System Failures} in the components responsible for orchestrating task execution, as well as issues related to dependencies between tasks in the Framework Layer.
These failures can significantly impact task scheduling, monitoring, and execution.
Examples of system failures include \textit{Monitor Loss}, where the component responsible for overseeing the execution of tasks and maintaining the state information becomes unavailable or unresponsive; \textit{Manager Loss}, where the component responsible for managing task scheduling, resource allocation, and overall workload coordination fails; and \textit{Dependency Failures}, where the frameworks fail to manage dependencies between tasks. 

Failures at the \textbf{Runtime Layer} are \textbf{Resource Failures} that occur while managing and using computational resources required to execute tasks. 
These errors can affect the availability, allocation, and effective utilization of resources, leading to disruptions in task execution.
Examples include \textit{Resource Starvation} where tasks do not receive sufficient resources (CPU, memory, storage, etc.) for execution, and \textit{Pilot Job Initialization Failure} where the pilot job responsible for provisioning and managing computational resources fails to start or initialize correctly.

Failures in the \textbf{Environment Layer} are 
\textbf{Hardware \& Environment Failures} related to the physical infrastructure and the overall runtime environment in which tasks are executed. 
Hardware failures are particularly common.
% , reaching 58\% in cloud computing \cite{Taxonomy} and 64\% \cite{failure_study}, 42.1\% \cite{BlueWaters} in HPC. 
% analyzed the number of 
For example, 42.1\% of all failures observed in the Blue Waters supercomputer over 261 days in 2013 were hardware failures~\cite{BlueWaters}, as were 64\% of failures seen in 22 HPC systems over nine years (1996-2005)~\cite{failure_study}.
Such errors can cause significant disruptions, including application or complete system failures.
% incompatibilities that prevent tasks from running correctly.
Examples include \textit{Hardware Shutdown}, where components such as servers, storage devices, or network equipment unexpectedly power down or fail, and \textit{Runtime Environment Mismatch}, where the software environment required for executing tasks does not match that available on execution nodes.

\subsection{Failure Manifestation} \label{subsec:manifestation}
% \sicheng{This part is similar to the next section (detecting failures). What's the difference between these two parts? How can we modify them?}
% \haochen{This section could account for a few examples you have at hand, about how exception messages are returned in the stack trace (e.g., error message E originated in class X, wrapped through class Y, transmitted through object Z, manifest at log at method A) and how lost of service heartbeats are detected (e.g., after x seconds of no response, loop X exits and sends an error message to user through...) You can discuss how failures and errors are represented by the TBPP systems, so that the next section can discuss how you would capture these signals. } 
% \sicheng{Does the current content here look good?}

Failure manifestation refers to the observable signs or indicators that a failure has occurred. These manifestations help in identifying, diagnosing, and addressing errors. They can appear in various forms, such as exception messages, service heartbeats loss, resource usage logs, and other log messages.

\textbf{Exception Messages} are error messages generated by the system when it encounters an unexpected condition or error. These messages typically provide information about the nature of the error and its location in the code. Examples are SyntaxError, FileNotFoundException, and OutOfMemoryError.

A \textbf{Service Heartbeat} is a periodic signal sent by a service to indicate that it is operational. The absence of a heartbeat can indicate that the service has failed or become unresponsive. 
%which helps detect System Failures.

\textbf{Resource Usage Logs} track the utilization of system resources such as CPU, memory, disk, and network. Abnormal patterns in resource usage can indicate failures or performance issues. 
%This is useful in detecting resource failures. 
For example, a steady increase in memory usage without release may indicate a memory leak, and a spike in CPU usage may indicate resource contention and a potential bottleneck.

\textbf{Other Log Messages} generated by the system can also provide insights into failures. TBPP frameworks have different levels of log messages, including debug information, warnings, errors, and informational messages. Feedback from submitted jobs (i.e., standard output and error files) also contains useful information for failure detection.

\section{Monitoring across Task Hierarchicies} \label{sec:detect}

In this section, we describe the monitoring system of \name{} and describe how it can detect failures.

% \subsection{Hierarchical Monitoring System} \label{subsec:monitoring}

As we have described, failures may occur at different layers of TBPP frameworks.  We therefore implement \name{} with a hierarchical monitoring system to observe the execution of tasks in TBPP frameworks.
The hierarchical monitoring system consists of \emph{task monitoring agents}, \emph{system monitoring agents}, \emph{a centralized monitoring database}, and \emph{a communication radio}.

\textbf{Task monitoring agents} are processes distributed across the system to observe the behavior of individual tasks at different levels of the task hierarchy. 
For example, an agent at each compute node is responsible for monitoring real-time metrics such as CPU, memory usage, and execution time, and an agent at the central TBPP manager collects information such as task metadata, task dependencies, and task states (e.g., submitted, ready, completed).

\textbf{System monitoring agents} focus on monitoring node failures and resource availability. Specifically, these agents running on hardware nodes periodically send \emph{heartbeats} to agents at a higher level and the centralized database. If a heartbeat is missed or delayed several times, the system identifies the node as potentially failed or overloaded.

The \textbf{centralized monitoring database} consolidates data from task-level monitoring agents, enabling efficient retrieval, analysis, and decision-making. This database simplifies data access for TBPP frameworks, facilitating quick retrieval of monitoring information and supporting effective failure detection and subsequent recovery actions.

The \textbf{communication radio} enables communication between the monitoring agents and the centralized database and is responsible for relaying failure alerts, status updates, and metrics between different layers of TBPP frameworks.

\section{Resilience Module} \label{sec:respond}

The resilience module in \name{} relies on a \emph{failure categorization engine} and a \emph{policy engine} to handle failures when failures are detected. 

\subsection{Failure Categorization Engine}
The failure categorization engine is responsible for analyzing detected failures and categorizing them based on predefined rules and historical data. Different types of failures, such as hardware faults, software bugs, or resource issues, may require different responses. 

The failure categorization engine maintains a \textbf{failure taxonomy library} of failure types and associated recovery mechanisms, which can be referenced to determine the best response. 
Specifically, for failures that occur at the application layer, we summarize the exceptions and errors that may occur in Python. %~\cite{python_error}.
Other failure types are recorded based on the categorization methods introduced in \S\ref{sec:characterize}, which is done by the failure root cause analyzer.

The \textbf{failure root cause analyzer} in this engine performs a comprehensive root cause analysis on the collected monitoring data from multiple layers (application, framework, runtime, and environment) and multiple sources (e.g., logs, exceptions, and resource logs) whenever a failure is detected.
This analyzer not only identifies different types of failures but also performs a resource analysis to determine if the failures are due to resource mismatches.
Based on the analyzed results from the failure root cause analyzer, \name{} selects an appropriate recovery mechanism using the \emph{resilience policy engine}, detailed in the next subsection.

\subsection{Resilience Policy Engine}
\begin{figure}[h]
    \centering
    \includegraphics[page=1, width=\linewidth, clip=true, viewport=0 00 565 400]{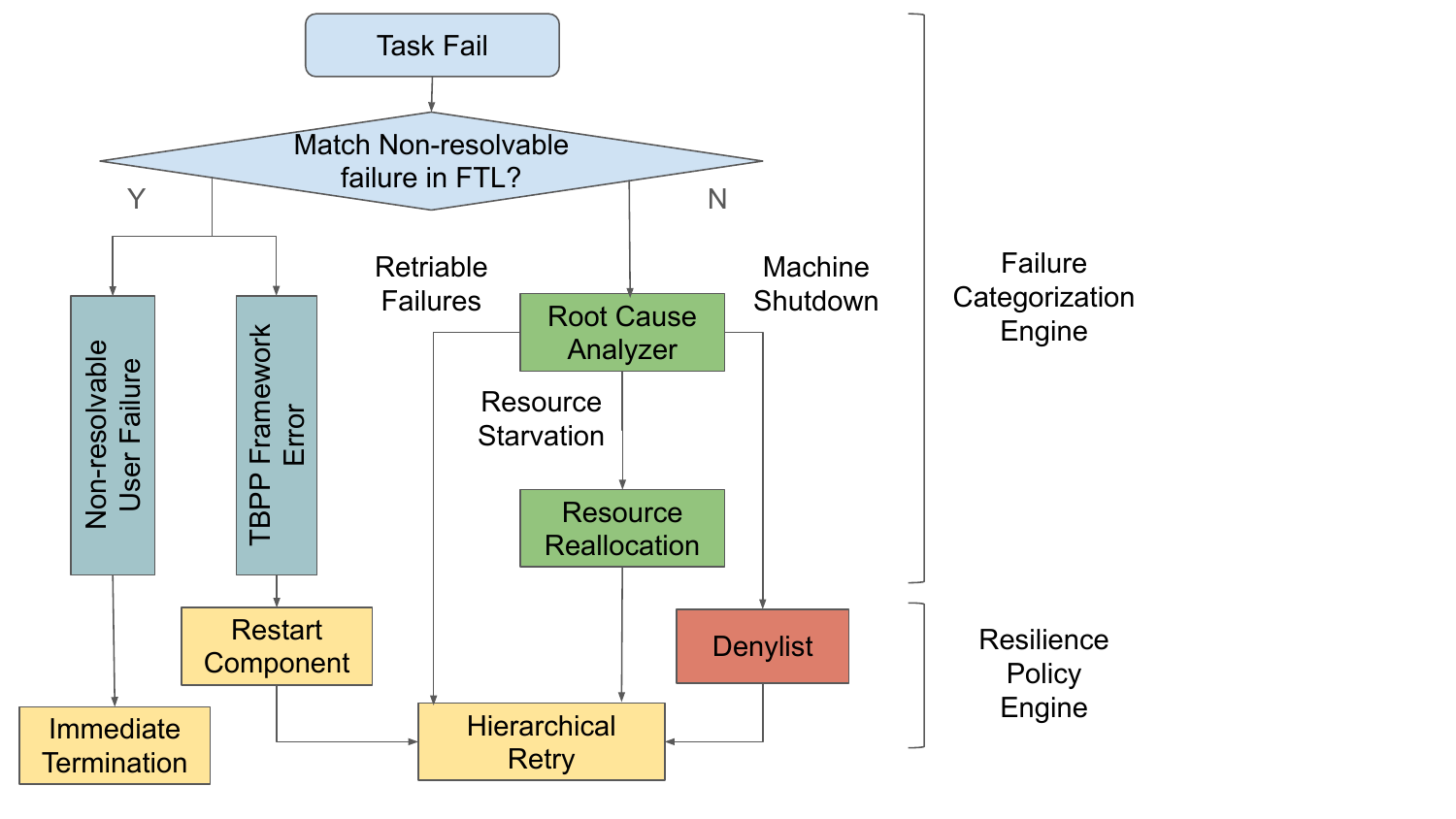}
    \caption{Flow of the failure categorization engine and resilience policy engine. FTL: Failure Taxonomy Library. 
    %\zhuozhao{font black}
    }
    \label{fig:resilience}
     \vspace{-1.5em}
\end{figure}

For each failure, the resilience policy engine provides an appropriate failure-handling strategy based on the categorizations provided by the failure categorization engine. It contains predefined policies: sets of rules and actions designed to address different failure scenarios at various layers.
Example action list in the policy engine includes \emph{resource denylist}, \emph{immediate termination}, \emph{hierarchical retry}, and  \emph{restarting system components}, as shown in \autoref{fig:resilience}.

The policy engine maintains a \textbf{resource denylist} which records the malfunctioning components in the TBPP framework. The system monitoring agents use the heartbeat mechanism to monitor the status of each component; those that fail to communicate are considered lost and are added to the resource denylist. 
As in HTCondor~\cite{condor}, resources could be removed from this list if they later resume communication.
%This ensures that the assigned resources are working normally.
%\ian{Could mention that, as in Condor, resources can move on/off the denylist based on monitoring.}

% \ian{Do we have data on how often the different mechanisms listed in the following are used, and how well they do?} \sicheng{you mean in literature?}
% \ian{I meant in the experiments that you have conducted}

When the failure root cause analyzer indicates that a failed task is non-recoverable, the policy engine maps it to an \textbf{immediate termination} action, which results in the termination of both the task and the application.
This decision is made to prevent further resource consumption since continuing to execute a non-recoverable task can lead to wasted computational resources.

The policy engine employs a \textbf{hierarchical retry} mechanism to reschedule failed tasks across the hierarchy of resource pools in TBPP frameworks. It attempts these steps in turn: 
%It by progressively adjusting the execution environment, starting with minimal changes and then escalating to more significant changes if the issue persists. 
%Specifically, the hierarchical retry would
\begin{enumerate}
    \item attempt to retry according to the resource requirements provided by the failure categorization engine to solve certain resource insufficiency problems;
    % on any resources first regardless of failure types;
    \item retry the task on a different node of the same resource pool in case the task requires specific execution environments;
    \item retry where the task has historically succeeded most frequently, ensuring an informed and adaptive retry process;
    \item retry on different resource pools (e.g., different Parsl executors~\cite{babuji2019parsl} or clusters/endpoints in Pegasus~\cite{Pegasus}).
\end{enumerate}

System failures in TBPP frameworks are often relevant to the failures of the framework components (e.g., the central manager or node managers). In such cases, the policy engine will first attempt to identify the failed components within the system. Once located, it will initiate a \textbf{restart of the failed components} to restore functionality. Following this recovery action, the policy engine can then perform a hierarchical retry of any affected tasks, reassigning them to available resources as needed. 

\subsection{Overall Failure Handling Flow}

The overall flow of the resilience module is illustrated in \autoref{fig:resilience}.
 The process is as follows:
 
\begin{enumerate}
    \item  
    The module first examines whether the failures are non-recoverable from hierarchical retries. If failures are deemed non-recoverable: 
    \begin{itemize} 
    \item For non-recoverable user failures, \name{} immediately terminates the execution. 
    \item For system failures within TBPP frameworks, \name{} attempts to restart the failed components and subsequently performs hierarchical retries. 
    \end{itemize}
    \item If the failures are recoverable, the resilience module utilizes the failure root cause analyzer to identify the specific failures. If these failures are not related to resource issues, \name{} proceeds to execute hierarchical retries directly.
    \item In cases where resource failures are identified, the resilience module analyzes the resource profile data to ascertain whether the failure is due to resource starvation or machine shutdown. Based on this analysis, \name{} provides tailored hierarchical retry suggestions to the most appropriate resource pools.
\end{enumerate}

\section{Implementation} \label{sec:implementation}

% \valerie{is it just me or do you only mention WRATH 2/3x in the paper? Is wrath just the resilience module or does it include the monitoring system. Maybe a system architecture diagram is required here} \sicheng{I think WRATH is the whole picture which includes the monitoring system and the resilience module. What kind of system architecture diagram do we want here?}\valerie{maybe something to show the flow of monitoring information from parsl to WRATH components (and also where are the hooks to wrath in parsl) and then the flow from WRATH back to parsl} \sicheng{I hope fig \ref{fig:WRATH} is what you want here.}
% \zhuozhao{describe general things (e.g., TaskRecord, conceptual model) in the previous sections, and describe specific implementation, like implemented in Parsl, use what library}
% \zhuozhao{We implement XXXX in Parsl, a typical Python-based task parallel }
% \kyle{Its not clear how this maps to the text in 4.}

We implement a prototype of \name{} and integrate it into Parsl~\cite{babuji2019parsl}, a widely used Python-based TBPP framework. The implementation includes about 3k+ lines of code and is open-source\footnote{\url{https://github.com/ClaudiaCumberbatch/resilient_compute}}.
All the components of \name{} are modularized and can be easily extended to support any appropriate alternatives.

%, a typical Python-based task parallel scripting library. 
% The code is available in \cite{parsl_resilience} and \cite{resilience_compute}.

\subsection{Parsl Introduction}

Parsl~\cite{babuji2019parsl} is a Python library for developing parallel and distributed programs. It provides a flexible and scalable runtime for executing scientific workloads and data-intensive applications across various computing resources. Here we introduce the Parsl architecture before describing how we implement \name{} in Parsl. 

\textbf{DataFlowKernel (DFK)} is the central manager responsible for managing the flow of tasks and data in the workload. Its functions include dependency resolution, i.e., analyzing the dependencies between tasks and controlling their execution order; task scheduling, i.e., submitting the task to an appropriate executor for execution; and task status tracking.

\textbf{Executors} define the type of computational resources and are responsible for distributing tasks to node managers. They maintain lists of active managers and schedule tasks to them based on their capacities. 

\textbf{Node Managers} are responsible for provisioning and allocating resources on an individual node. They ensure that the workers are properly launched and track their status via heartbeat messages. They also maintain task queues and result queues, from which workers pull tasks and to which workers push task results, respectively.

\textbf{Workers} are processes that execute the actual tasks in Parsl. They pull tasks from the task queue to run. Multiple workers can run in parallel on the same node, allowing for efficient utilization of available resources.

\subsection{\name{} over Parsl} \label{subsection:monitoring_system}
\begin{figure}[ht]
    \centering
    \includegraphics[page=1, width=\linewidth, clip=true, viewport=0 0 610 400]{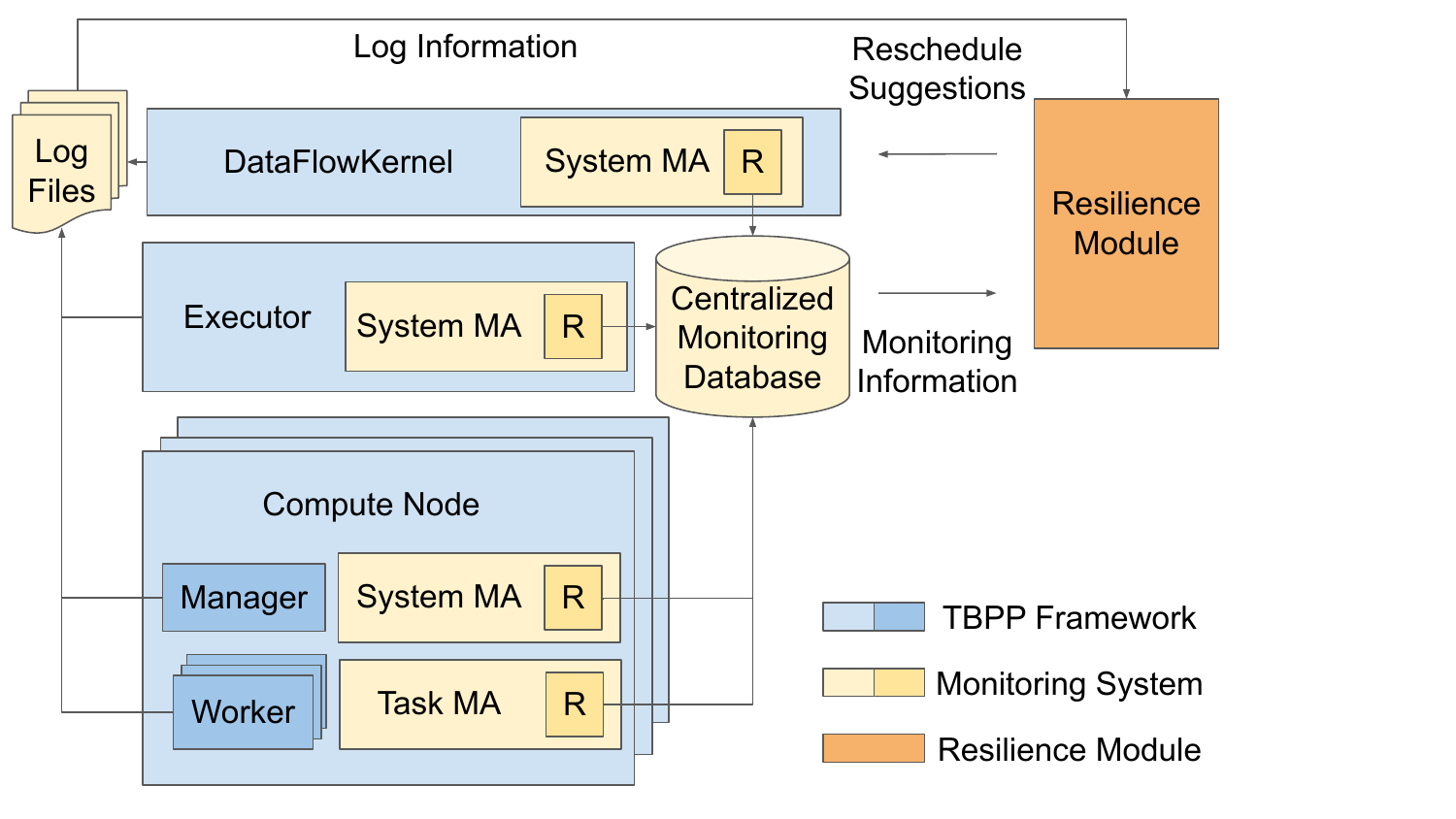}
    \caption{\name{} system architecture diagram. Components in yellow and orange denote components of \name{}. MA: Monitoring Agent. R: Communication Radio. 
    }
    \label{fig:WRATH}
     \vspace{0em}
\end{figure}

The overall architecture of \name{} over Parsl is shown in \autoref{fig:WRATH}.
We implement \name{}'s task monitoring agents across the hierarchy of Parsl. 
Specifically, for each node manager in Parsl, the monitoring system launches a \emph{node-level process} that employs Python's \texttt{psutil} library~\cite{psutil} to collect the resource information (e.g., CPU and memory utilization) of all the tasks running on that node.
\change{This resource profile data, along with detailed task information and the status of each Parsl component, is transmitted via the \emph{radio}, an interface based on the TCP protocol.
The radio operates across various locations in the system (e.g., workers, nodes, and the central manager) and sends monitoring data to a modular database.}
% The task information, including the status of each Parsl component and resource profile data, is sent by the \emph{radio} to a centralized database. 
% The \emph{radio} is implemented using the TCP protocol.
Currently, \name{} supports sending monitoring information to a local database, cloud-hosted database, 
or a cloud-hosted event fabric for scientific computing (Octopus~\cite{octopus}) to trigger later events.

We implement the resilience module as a \emph{retry handler} in Parsl. 
When a task fails, the Parsl DFK automatically invokes the retry handler to determine how to handle the failed task. 
\change{The failure root cause analyzer in \name{} uses a decision tree to classify errors. For unrecoverable errors, the analyzer combines task and system metrics with heuristics (e.g., error types and retry counts) to recommend fail-fast decisions.
To handle Python package-related failures, \name{} dynamically collects package availability on compute nodes (via \texttt{pip freeze}) and matches task requirements (using static analysis) to identify suitable nodes. Non-Python library or package-related failures are classified as application-layer errors and flagged as non-recoverable, requiring user intervention.}

\change{By adopting a layered architecture for failure characterization, hierarchical monitoring, and a resilience module, the core ideas of \name{} are framework-agnostic and accommodate diverse environments. Users can define custom rules for failure categorization and retry strategies, enabling seamless integration with existing monitoring and resource management systems. As outlined in Section 2, this approach is generalizable to most TBPP frameworks, given their layered architectures and the reliance on retries for failure recovery.}

\begin{table*}[ht]
\caption{The benchmark applications.}
\label{tab:applications}
\begin{tabularx}{\textwidth}{@{}lXrlX@{}}
\toprule
\textbf{Application} & \textbf{Description} & \textbf{\# Tasks} & \textbf{Configuration} \\ \midrule
Cholesky & Compute Cholesky decomposition of a randomly generated positive definite matrix & 385 & Matrix size: 10 000*10 000, Block Size: 1000*1000 \\ \midrule
Docking & Predict orientation and position of two molecules & 160 & Initial simulations: 8; Batch size: 8; Rounds: 3 \\ \midrule
FedLearn & Federated learning & 5 & Dataset: MNIST; Clients: 8; Batch size: 3, Rounds: 3, Epochs/Round: 3 \\ \midrule
MapReduce & Count words using a MapReduce strategy & 101 & Dataset: Randomly generated, Map task count: 100, Generated files: 100 \\ \midrule
MolDesign & Use ML to identify molecules with largest ionization energies & 214 & Initial simulations: 4; Batch size: 4; Search count: 16 \\ %\midrule
% montage & Takes a directory of input astronomy images and stitches them into a single mosaic using Montage tools. & 419 & -- \\ 
\bottomrule
\end{tabularx}
\end{table*}

\begin{table*}[ht]
\caption{Failure types that we inject during our \name{} experiments}
\label{tab:inject}
\begin{tabularx}{\textwidth}{@{}llXX@{}}
\toprule
\textbf{Layer (Failure Type)}                         & \textbf{Injected Failures} & \textbf{Description}               & \textbf{\name{} Solution}            \\ \midrule
\multirow{2}{*}{Application Layer (User Failure)}       & Zero-division              & Raise divide by zero error           & Terminate            \\ 
                                                        & Failure                    & Raise runtime exception          & Terminate                                 \\ \midrule
\multirow{2}{*}{Framework Layer (System Failure)} & Worker-killed              & Kill current process        & Reschedule to another worker                \\
                                                        % & manager-killed             & Kill the parent process (i.e., the manager).                                       \\
                                                        & Dependency                 & Parent task exception leads to child task dependency failure       & Act according to the root cause of the dependent item   \\ \midrule
\multirow{2}{*}{Runtime Layer (Resource Failure)} & ulimit & Open 1M files to simulate ulimit exceeded error   & Hierarchical retry  \\
                                                        & Memory                     & Force out of memory error   & Allocate sufficient memory                                         \\ \midrule
Environment Layer      & Import & Simulate import error due to bad environment  & Hierarchical retry \\ 
(Hardware \& Environment Failure) & & & \\
\bottomrule
\end{tabularx}
\end{table*}

\section{Evaluation} \label{sec:evaluation}

We evaluate \name{} by applying it to a number of Parsl applications and aim to answer the following questions. 

\begin{itemize}
    \item Can \name{} accurately identify non-resolvable failures, stop retrying, and "fail fast" to minimize wasted time and resources?
\item When resolvable errors occur, does \name{} enhance the application's success rate by making appropriate retry decisions?
\item Does \name{} introduce significant overhead to application performance?
\item How does \name{} perform as the failure rate and scale of the applications increase?
\end{itemize}

\subsection{Experimental Setup}

\textbf{Testbed:} 
We conduct our evaluation on an HPC cluster\footnote{\url{https://hpc.sustech.edu.cn/introduction/hardwareresource.html}}. The cluster consists of 815 CPU nodes (each with 2 Xeon Gold 6148 CPUs and 192 GB of memory) and 2 large-memory nodes (each with 8 Xeon Platinum 8160 CPUs and 6 TB of memory).

\textbf{Workloads:} 
We evaluate \name{}'s performance on five applications from the TaPS benchmark suite~\cite{pauloski2024taps}, which provides multiple real-world, DAG-based applications to benchmark the performance of TBPP frameworks: see \autoref{tab:applications}. We use Parsl as our TBPP framework here. 
% To emulate task failures, we implement a failure-injection application in TaPS. 
% In normal applications, tasks are submitted by \emph{engines}, which support different types of TBPP frameworks. 
% \valerie{Are you referring to the, e.g., HighThroughputExector here? I think it's unclear what engine is. Maybe ``task scheduler" instead? Or i guess you're talking about the TaPS definition of engine. Maybe just say that you've implemented a failure ingection engine that replaces normal tasks with failure tasks, and remove the sentence above about normal applications}
% \sicheng{yes I mean TaPS engine here. I feel like describing the normal application will help the reader to understand what engine is here?} \valerie{I don't think you describe TaPS much, so i don't think readers will know what a TaPS engine is, just replace with task-scheduler or something of the sort}
%\ian{To be clearer here, you could say that for these Parsl examples, you normally have a "Parsl engine" (?) and you replace this with a "Parsl-fail engine" that replaces F\% of each tasks that it is ask to execute with a failure task (?).}
%\sicheng{Thank you Ian, it is much clearer to explain like this. I have updated it. Please check.}
To emulate failures, we created \emph{failure-injection engines} in TaPS that replace selected tasks with failure tasks. 
% A failure task is a task that fails with a given failure type according to the specified failure rate. 
%during submission.
For example, we replace the standard Parsl engine with a ``Parsl-fail engine''. This modified engine allows us to replace a specified fraction of the tasks in the benchmark applications with a failure task. 
%, which wraps the original execution engine to create a failure-injection engine, which is initialized with a given failure rate and a failure type and will create failure tasks according to these two parameters during runtime. 
The supported failure types are described in \autoref{tab:inject}. 
In the experiments that we describe below, we use this machinery to evaluate the performance of each of our five applications as we vary the types of failure.

\begin{figure*}[ht]
        \centering
        \includegraphics[width=\textwidth]{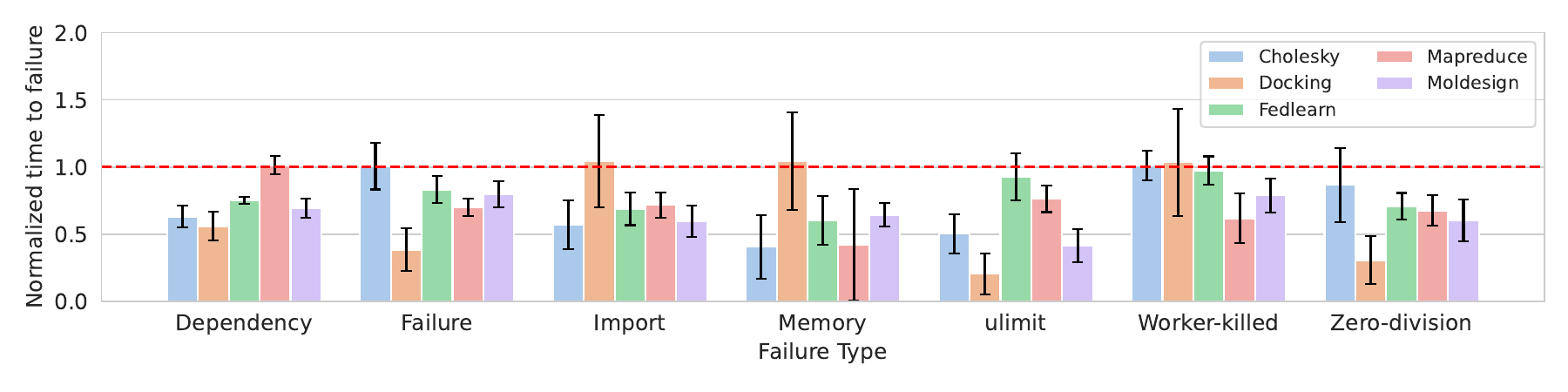} 
        \caption{Normalized time to failure for the applications with different failure types when \name{} is enabled. All results are normalized to those without \name{}. Failure rate = 0.3, Nodes = 32. 
                Error bars represent the standard error of the mean (SEM) across 10 independent runs. All of the trials failed here, but those with \name{} failed fast. 
                }
        % \label{fig:makespan}

    \label{fig:makespan_and_overhead}
\end{figure*}

\textbf{Metrics:} We evaluate the effectiveness of \name{} using the following metrics. 

\begin{itemize}
    \item Makespan: The total time taken to complete all tasks of an application, including TBPP framework initialization, task execution, retry, clean up, etc. 
    %\item Time to success: The makespan to finally successfully run an application, including the time to retry.
    \item Time to failure: The makespan at the point when any task in the application fails without remaining retry attempts, resulting in the application's failure.
    % \item Overhead per task: The time spent by \name{} to categorize failures and make decisions on how to retry for each task, on average.
    \change{\item Overhead ratio: The proportion of time consumed by \name{} to analyze failures and decide retry strategies, relative to the total makespan of the workflow.}
    % \item Average task time: Total time used by all tasks divided by the number of trials of tasks.
    \item Task success rate: The number of successful tasks divided by the total number of tasks.
    \item Retry success rate: The number of successfully retried tasks divided by the total number of tasks that were retried.
    % \zhuozhao{add application success rate}
    \item Application success rate: The percentage of runs that are successfully completed without any failures during the application execution across multiple runs. 
   % \item Worker utilization \sicheng{still thinking of how to calculate this}
\end{itemize}

\textbf{Baseline:} We use Parsl and its default retry mechanism as the baseline for comparison. In this default mechanism, tasks are always retried on the same Parsl executor, regardless of the failure type or resource availability. 

% The real-world applications include Cholesky decomposition, protein Docking, FedLearn, MapReduce, molecular design, and montage. Some general descriptions and configurations for our experiments are provided in Table~\ref{tab:applications}. 

\subsection{Overall Performance of \name{}}

% \sicheng{Not sure whether to keep this subsection.}
% In this subsection, we evaluate the performance of \name{} on different applications with sporadic errors. 
% We ran different combinations of base applications, failure types, and with or without \name{} on two nodes. The applications supported by TaPS and the corresponding configurations we used in this experiment are described in Table~\ref{tab:applications}. The failure rate is set to 0.3 which means each submitted task will have a 30\% of possibility to run as an intentional failure function instead of the application task. 
% We ran each combination for ten trials and then calculated the normalized makespan, which is the makespan with \name{} divided by the makespan without \name{}, and the overhead, which is the time spent on invoking the resilience module.

% We measured makespan and overhead as evaluation criteria. Makespan refers to the total time taken to complete all tasks in a given workload, including TBPP framework initialization, application execution, and shutdown. Overhead quantifies the additional computational and time costs introduced by the resilience module itself.
% For a certain configuration of an application, a lower makespan indicates better performance and a lower overhead indicates a more efficient resilience module that minimizes its impact on overall system performance. 

\textbf{Time to failure:} In this experiment, we run the benchmark applications on 32 nodes. To stress-test \name{}, we set the failure rate as 0.3, meaning that 30\% of tasks in each application are replaced with failure tasks. 
\autoref{fig:makespan_and_overhead} shows the time to failure
for different applications across various failure types with \name{} enabled. %Results are normalized to the makespan without \name{}.
The results show that applications tend to fail more quickly with \name{} compared to without it. 
For most application and failure type combinations, \name{} reduces the time to failure by 20\%–50\%. This is because \name{} identifies the root cause of failures and makes more informed retry decisions, allowing tasks that are destined to fail to do so more rapidly.
%either letting the tasks fail quickly or allocating them to resources that are more likely to accomplish the tasks. %, thus reducing the makespan. 
% For user failures and environment failures like import and zero-division, workloads need only to spend about 1/3 of the time to complete, because the resilience module can identify those errors and reduce the need for redundant retries. For system failures, the strategy of hierarchical retry also improves efficiency. When it comes to resource failures, the resource reallocation strategy guarantees that tasks avoid the previous failure causes. 
% We can also observe that Docking and FedLearn did not perform well in some of the failure types. That is because they have relatively long execution times, resulting in a large amount of profiling data for the resilience module to analyze. But this happens only to a small number of failure types. 

% \begin{figure}
%     \centering
%     \subfloat[Success rate for import failure.]{\includegraphics[width=0.4\linewidth]{Images/resolvable_import.pdf}}
%     \subfloat[Success rate for memory failure.]{\includegraphics[width=0.4\linewidth]{Images/resolvable_memory.pdf}}
%     \caption{Task success rate and retry success rate of MapReduce.}
%     \label{fig:resolvable}
% \end{figure}

However, the error bars indicate significant variability in our experimental results. This is because failures can be injected at any point within the application's DAG, and different failure points can lead to substantial differences in application behavior. Take a MapReduce application as an example. a failure can occur during either the Map stage or the Reduce stage. If a failure occurs in the Map stage, the Reduce stage is not executed due to unmet dependencies. In contrast, a failure in the Reduce stage occurs after substantial time has already been spent executing tasks in the Map stage, resulting in a longer time to failure. For applications with more complex DAG structures, the variability between runs becomes even more pronounced. 

% \valerie{I think we should flag which errors are resolvable and which ones aren't. It's too bad \name{} ``failed-fast" so often on memory errors, which are otherwise resolvable (unless they weren't in this experiment?). I think the discussion on the error bars explains some but not enough} \zhuozhao{I think Table III show which failures are resolvable}\valerie{Yes, but it doesn't explain why you'd want them to fail fast here nor why they haven't succeeded}
% \zhuozhao{it seems the results do not show much on this..}
% \valerie{I'm going to comment out these messages, but I feel like these are the questions I'd be asking myself if i were going through the results.} \zhuozhao{I wonder that too}

% \textbf{Overhead:} \autoref{fig:success_overhead} illustrates the overhead per task for successful runs. In all experiments, the overhead per task is less than 0.05~s, and in most cases, it is under 0.03~s.

\begin{figure}
    \centering
    \includegraphics[page=1, width=0.8\linewidth]{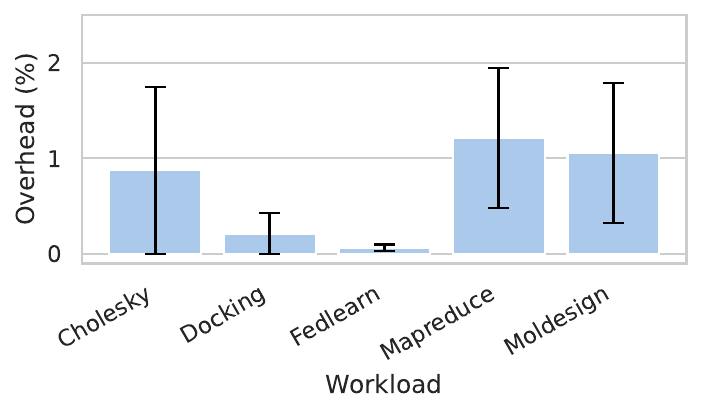}
    \caption{
    % Average overhead of \name{} on successful runs of each application with a pre-set failure rate of 0.3 on 32 nodes.
    \change{Overhead ratio of \name{} on successful runs of each application with a pre-set failure rate of 0.1 on 32 nodes.}
    }
    \label{fig:success_overhead}
     \vspace{0em}
\end{figure}

\change{\textbf{Overhead:} \autoref{fig:success_overhead} illustrates the overhead ratio for successful runs. In all experiments, the overhead ratio is less than 2\%, and in most cases, less than 1\%.}

\subsection{\name{}'s Performance for Resolvable Failures}\label{sec:resolvable}

In this experiment, we aim to show how \name{} performs hierarchical retries for the MapReduce application when dealing with two types of resolvable failures, i.e., memory-insufficient failures and import failures. Note that similar results can be obtained for all the other applications, so we only show the results of MapReduce due to space limits.
The settings for the two failure types are as follows:
\begin{itemize}
    \item \textbf{Memory failure}: Each task requires 200~GB of memory and runs on a single node. We configure two executors in Parsl: one with nodes that have 192~GB of memory and another with nodes that have 6~TB of memory. 
    \item \textbf{Import failure}: Each task requires a specific software package. We configure two executors in Parsl: one that has the required package installed and one that does not. 
\end{itemize}

% \begin{figure}
%     \centering
%     \subfloat[Results for import failures. X axis is the number of nodes without the required package. The number of nodes with the required package is fixed to one.
%     ]{\includegraphics[width=0.8\linewidth]{Images/scale_import.pdf}}
    
%     \subfloat[Results for memory failures. X axis is the number of nodes with insufficient memory. The number of nodes with sufficient memory is fixed to one. ]{\includegraphics[width=0.8\linewidth]{Images/scale_memory.pdf}}
%     \caption{Application success rate of MapReduce when being injected with different types of failures.}
%     \label{fig:scale}
% \end{figure}

\autoref{tab:resolvable} shows that \name{} significantly improves both the task success rate and the retry success rate. 
This is because the tasks can only succeed when they are allocated to (or retried on) the appropriate executor---either the one with sufficient memory or the one with the necessary package.
Without \name{}, tasks are repeatedly retried on the same executor, meaning that if a task fails due to a memory or import failure, it will continuously fail. Thus, it's unsurprising that both the task success rate and retry success rate are low in the absence of \name{}. With \name{}, tasks are retried across different resource pools based on the success rate and resource availability of each executor, leading to a higher success rate during execution.

\begin{table}[h]
\centering
\caption{Task success rate and retry success rate of MapReduce. SR: Success Rate.}
\label{tab:resolvable}
\begin{tabular}{@{}cccc@{}}
\toprule
\textbf{Configuration}                     & \textbf{Failure Type} & \textbf{Retry SR} & \textbf{Task SR} \\ \midrule
\multirow{2}{*}{Parsl with WRATH} & import       & 0.53               & 0.43              \\
                                  & memory       & 0.75               & 0.47              \\ \midrule 
\multirow{2}{*}{Parsl w/o WRATH}  & import       & 0.22               & 0.00              \\
                                  & memory       & 0.24               & 0.00              \\ \midrule 
\end{tabular}
\end{table}

% \begin{figure}
%     \centering
%     \subfloat[Success rate for import failure.]{\includegraphics[width=\linewidth]{Images/resolvable_import.pdf}}
    
%     \subfloat[Success rate for memory failure. 
%     % \sicheng{this is only 10 trails, more trails are in progress}\zhuozhao{enough for now}
%     ]{\includegraphics[width=\linewidth]{Images/resolvable_memory.pdf}}

%     \caption{Task success rate and retry success rate of MapReduce.}
%     \label{fig:resolvable}
% \end{figure}

\subsection{Scalability}

\begin{figure}[ht]
    \centering
    \begin{subfigure}[b]{0.48\linewidth} 
        \centering
        \includegraphics[width=\linewidth]{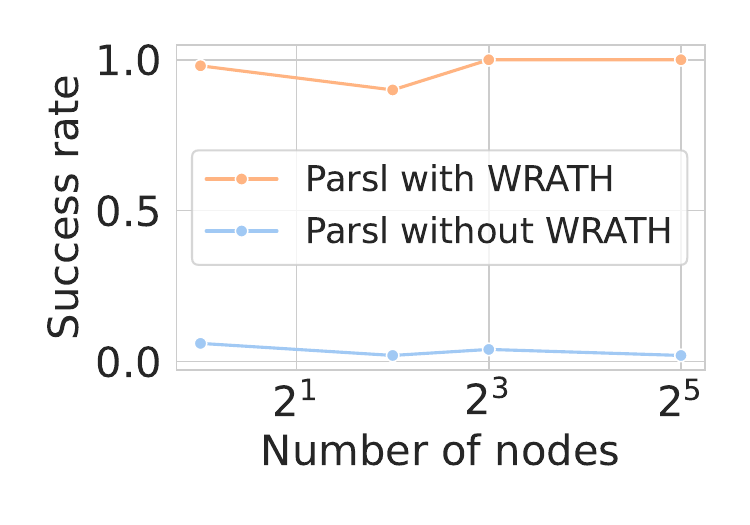}
        \caption{Results for import failures. X axis is the number of nodes without the required package. The number of nodes with the required package is fixed to one.}
        \label{fig:scale_import}
    \end{subfigure}
    \hfill 
    \begin{subfigure}[b]{0.48\linewidth}
        \centering
        \includegraphics[width=\linewidth]{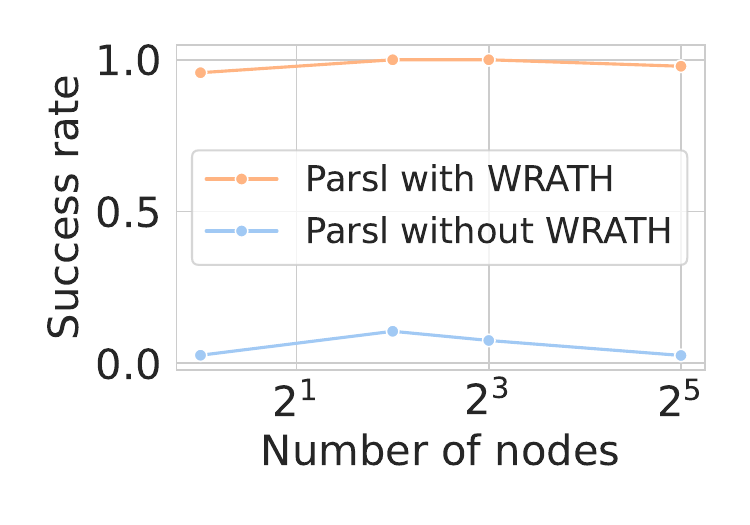}
        \caption{Results for memory failures. X axis is the number of nodes with insufficient memory. The number of nodes with sufficient memory is fixed to one.}
        \label{fig:scale_memory}
    \end{subfigure}
    
    \caption{Application success rate of MapReduce when being injected with different types of failures.}
    \label{fig:scale}
\end{figure}

In this experiment, we evaluate how \name{} performs when scaling the number of nodes. We inject either import failures or memory failures into the MapReduce application, using the same settings as those in \autoref{sec:resolvable}. We increase the number of nodes that either lack sufficient memory or do not have the required package.
As shown in \autoref{fig:scale}, \name{} consistently maintains an application success rate exceeding 90\%, regardless of the number of nodes with insufficient memory or missing packages. In contrast, without \name{}, tasks continuously fail. 
This enhancement is due to \name{}'s ability to leverage hierarchical retries, enabling it to effectively identify and allocate tasks to the appropriate resources for successful execution.

% \begin{figure}
%     \subfloat[Failure type = Import. X axis is the number of nodes without the required package. The number of nodes with the required package is fixed to one.
%     % \zhuozhao{y axis success rate}
%     ]{\includegraphics[width=\linewidth]{Images/scale_import.pdf}}
    
%     \subfloat[Failure type = Memory. X axis is the number of nodes with insufficient memory. The number of nodes with sufficient memory is fixed to one. ]{\includegraphics[width=\linewidth]{Images/scale_memory.pdf}}
%     \caption{Application success rate of MapReduce when being injected with different types of failures.}
%     \label{fig:scale}
% \end{figure}

% \autoref{fig:scale_overhead} shows that the overhead per task remains relatively constant as the number of nodes increases for both types of failures. The primary source of overhead is resource log analysis, as \name{} needs to process more logs when the number of nodes grows. This demonstrates that \name{}'s mechanisms for detecting failures and reallocating tasks are both efficient and scalable, with most of the overhead attributed to log processing rather than the failure handling or task redistribution itself.

\begin{figure}[h]
     \vspace{-1.5em}
    \centering
    \includegraphics[page=1, width=0.8\linewidth]{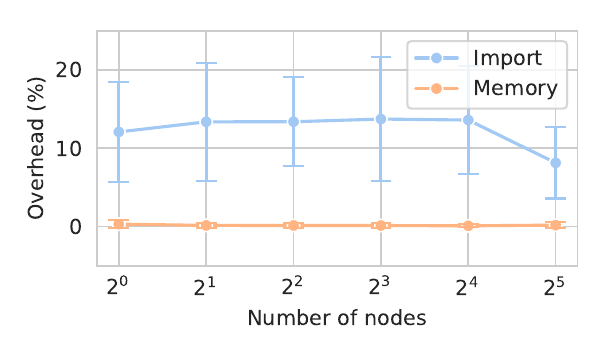}\vspace{-0.1in}
    \caption{
    % Overhead per task of \name{} with a varying number of nodes for the MapReduce application.
    \change{Overhead ratio of \name{} with a varying number of nodes for the MapReduce application.}
    }
    \label{fig:scale_overhead}
     \vspace{-1.5em}
\end{figure}

\change{\autoref{fig:scale_overhead} shows that the overhead ratio remains relatively constant as the number of nodes increases for both types of failures. The primary source of overhead is resource log analysis, as \name{} needs to process more logs when the number of nodes grows. This demonstrates that \name{}'s mechanisms for detecting failures and reallocating tasks are both efficient and scalable, with most of the overhead attributed to log processing rather than the failure handling or task redistribution itself.}

% \begin{figure}
%     \centering
%     \includegraphics[page=1, width=\linewidth]{Images/mapreduce_overhead.pdf}
%     \caption{Overhead per task of \name{} with a varying number of nodes for the MapReduce application.
%     }
%     \label{fig:scale_overhead}
%      \vspace{-1.5em}
% \end{figure}

\subsection{Varying the Failure Rate}

\begin{figure}
    \centering
    \includegraphics[page=1, width=0.8\linewidth]{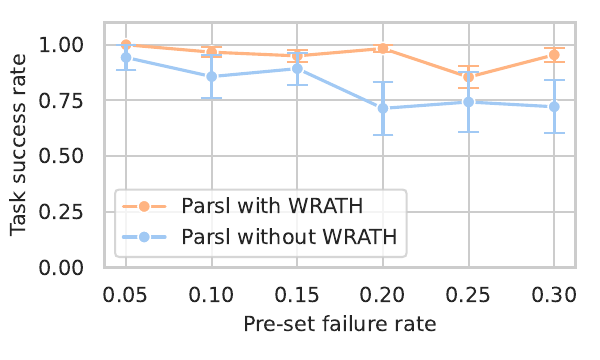}
    \caption{Task success rate of Cholesky when being injected different rates of memory failure. The number of small memory nodes is 16 and the number of large memory nodes is 1. Error bars represent the standard error of the mean (SEM) across ten independent runs. 
    }
    \label{fig:cholesky_rate}
     \vspace{-1.5em}
\end{figure}

In this experiment, we vary the failure rate within the range of $[0.1, 0.3]$ to evaluate how \name{} performs. 
We use the Cholesky application, randomly replacing tasks with a pre-set failure rate for memory-intensive tasks that require 200 GB of memory.
All other settings remain consistent with those in \autoref{sec:resolvable}. 
\autoref{fig:cholesky_rate} shows that \name{} maintains a high task success rate due to the hierarchical retry mechanism, which monitors the resource usage of failed tasks and schedules them to the appropriate resources for successful execution. In contrast, the task success rate without \name{} continuously decreases as the failure rate rises, demonstrating the effectiveness of \name{} in handling increasing failure rates.

\section{Related Work} \label{sec:related_work}

\subsection{Failure Categorization}

Failures in HPC systems have been widely studied at various granularities.
Schroeder et al.\ analyzed a decade of failure data collected from Los Alamos National Laboratory (LANL) clusters and categorized failures into hardware, software, environmental, network, and human error categories while excluding user application issues~\cite{schroeder_understanding_2007, failure_study, el-sayed_reading_2013}.
In contrast, we consider user errors to be a significant failure source, and our categorization methodology is more refined, targeting the TBPP stack and decomposing the software layer into Framework and Runtime layers.  

Di Martino et al.~\cite{BlueWaters} provided a similar failure categorization for Blue Waters, a Cray supercomputer, highlighting how different failures impacted workload executions, from interrupting with failover to multiple node failures.
They found that only 2\% of missing compute node heartbeats were due to actual node problems; in most cases, heartbeats resumed without operator intervention. Other researchers~\cite{das2021systemic} have also noted a lack of clear evidence linking health faults to node faults. \name{}'s resilience module also treats service heartbeat as a key indicator of errors, but we avoid relying solely on this signal to inform our resilience strategies. 

Several papers~\cite{HPC_failure_study, gupta2017failures}
%, liu_large-scale_2018, rojas_analyzing_2019, das_systemic_2021} 
consider more refined failure categories that align with the nature of their data sources, such as job scheduling logs and Reliability, Availability, and Serviceability (RAS) logs. They observe that many failures (31--99\%) are due to user and application behavior, such as coding bugs, incorrect configurations, operational errors, and memory exhaustion. Our resilient module mitigates such user failures through robust error matching and automated recovery strategies. For recoverable failures, such as missing dependencies, the module informs the scheduler to initiate a hierarchical retry. In the case of syntax errors, it terminates the execution without triggering unnecessary retries.

\subsection{Failure Detection}

Detecting failures across multiple layers in HPC and TBPP systems remains a challenge. Wintermute offers an online Operational Data Analytics (ODA) framework to monitor system metrics such as CPU cycles and power usage~\cite{netti_dcdb_2020}. \name{} monitors more than numerical readings, leveraging system logs at multiple levels (application, workload, runtime) to identify failures.  Furthermore, \name{} incorporates resilient strategies that allow for best-effort recovery without operator intervention.

Huang et al.\ explored minimally invasive fault detection in MPI applications using side-band network and independent hardware coress~\cite{huang_investigating_2010}. While similar in goal, \name{}'s cloud-based event fabric (Octopus) can be configured to be used for monitoring, enhancing reliability by reducing dependence on local hardware.

Other monitoring systems 
%by Jha et al., Shaykhislamov et al., Tuncer et al., and Aksar et al. 
focus on detecting performance degradation~\cite{jha_holistic_2017}
% ,shaykhislamov_approach_2018,tuncer_online_2019,aksar_e2ewatch_2021}
, using time series data of resource usage and numerical metrics such as CPU, memory, I/O, and network utilization, alongside with systems and hardware error logs. These systems apply statistical and machine learning methods to compare and contrast data from normal and abnormal runs to identify anomaly signatures. However, these approaches are not directly applicable to our lightweight resilient module, which neither relies on historical data nor requires significant computational power for model training or fine-tuning. Instead, it focuses on addressing categorized failures in the TBPP workload by identifying recoverable failures and performing hierarchical retries where feasible.

% Our work differs by targeting failures beyond performance anomalies within the TBPP workload.
% \ian{Out of context here, but one approach to reducing effect of failures, if you can analyze the DAG, is to replicate tasks that have many dependencies. The work in \cite{malewicz2007tool} could be relevant.}

\subsection{Failure Handling}

% Failure handling techniques in distributed systems and HPC can be categorized as reactive~\cite{iamnitchi2000problem,zhao2016elastic, replication, retry, resubmission, custom, n-version}, proactive~\cite{Rejuvenation, self-healing, reemptive, prediction, monitoring}, or resilient~\cite{RL2015, NN}. 
Failure handling techniques in distributed systems and HPC can be categorized as reactive~\cite{iamnitchi2000problem}, proactive~\cite{Rejuvenation}, or resilient~\cite{RL2015}. 
Reactive methods employ techniques like replication, checkpoints, and retry to mitigate the impact of failures, particularly for long-running tasks, but repeatedly resubmitting tasks without addressing the underlying cause of failure will lead to a huge waste of resources.
% However, replication and checkpointing increase execution costs, especially for shorter tasks. Moreover, retrying tasks may lead to inefficiencies if the same task is resubmitted repeatedly without addressing the underlying cause of failure.
Proactive methods monitor a system and make predictions to maximize availability, assuming accurate fault predictions. While they aim to prevent failures through early detection, their effectiveness hinges on the accuracy of fault predictions. 
% Inaccurate predictions can lead to unnecessary overhead from executing recovery strategies that may not be needed. Additionally, offline predictive models may struggle to adapt to dynamic failure scenarios, resulting in a lack of responsiveness when unexpected issues arise.
Resilience methods leverage machine learning or adaptive learning models to recover from faults quickly by continuously interacting with the environment, but will become inaccurate when encountering a rapidly changing environment. 
% However, these approaches also face limitations; frequent model updates require substantial computational resources and data. If the environment changes too rapidly or unpredictably, models may become inaccurate or even fail to provide timely recovery solutions.
Apart from these, Zhang et al.~\cite{zhang2020trua}  proposed Trua, which handles failures by employing a historical failure data-based task replication strategy and using anomaly detection to filter out unusual failures.

We focus on reactive strategies such as resource reallocation and hierarchical retries. Traditional methods~\cite{zhao2016elastic, Shrink_or_Substitute_2018} often conduct checkpointing and replication before retry mechanisms, while our approach explores work placement as an alternative mitigation strategy for atomic tasks. Implementing this approach requires extensive coordination across TBPP layers and the monitoring system, making our work a unique contribution to failure handling in HPC. Kola et al.~\cite{kola2005faults} discussed silent failures in distributed systems—failures that either produce incorrect results without any error status or cause processes to hang. In contrast, we focus on failures that generate exceptions across various layers of the TBPP framework.

% \ian{Out of place reference: Kola et al.~\cite{kola2005faults} discuss silent failures and ways to deal with them. Also potentially relevant: \cite{das2021systemic} and/or \cite{das2020aarohi}.}

% Mukwevho et al.~\cite{mukwevho2018toward} categorize fault management approaches as reactive methods (RAMs)~\cite{iamnitchi2000problem,zhao2016elastic, replication, retry, resubmission, custom, n-version}, proactive methods (PRMs)~\cite{Rejuvenation, self-healing, reemptive, prediction, monitoring, das2020aarohi}, and resilient methods (RSMs)~\cite{RL2015, NN}.

\section{Conclusion} \label{sec:conclusion}

% One of the future works is to apply the resilience monitoring system and retry module in other WMSs.

We have proposed \name{}, an approach for detecting and handling failures in distributed TBPP. We surveyed common failure-handling mechanisms, including reactive, proactive, and resilient methods, and identified their limitations in dynamic and large-scale parallel systems. To improve TBPP robustness, we developed \name{} with a scalable monitoring system and an intelligent resilient module. Together these components detect, report, and reschedule failed tasks, leading to a reduction in makespan and better resource utilization. Our evaluation results demonstrate that \name{} is effective in enhancing TBPP frameworks by offering automatic failure recovery, improving task execution efficiency, and minimizing performance overhead.
However, \name{} currently has limitations in supporting compiled languages. These limitations stem from fundamental differences in error handling between interpreted and compiled languages. In future work, we plan to extend \name{} to support compiled languages by developing language-specific recovery mechanisms and exploring cross-language interoperability for heterogeneous environments. Additionally, we will investigate its integration with other distributed computing frameworks and scalability in large-scale systems.
% In future work, we plan to explore the integration of \name{} with other distributed computing frameworks and investigate its scalability in large-scale systems. 


\section*{Acknowledgment}
% This work was supported in part by the National Natural Science Foundation of China Grant No. 62202216, the Guangdong Basic and Applied Basic Research Foundation Grant No. 2023A1515010244, and the Shenzhen Science and Technology Program Grant 20231121101752002. This work was also supported by Center for Computational Science and Engineering at Southern University of Science and Technology.

% This work was partially supported by NSF awards 2209919 and 2004894, the Diaspora project, and by the U.S. Department of Energy (DOE), Office of Science, Office of Advanced Scientific Computing Research, under Contract DE-AC02-06CH11357.

This work was supported in part by the National Natural Science Foundation of China Grant No. 62202216, the Guangdong Basic and Applied Basic Research Foundation Grant No. 2023A1515010244, the Shenzhen Science and Technology Program Grant 20231121101752002, NSF awards 2209919 and 2004894, and the Diaspora project funded by the U.S. Department of Energy under Contract DE-AC02-06CH11357. This work was also supported by Center for Computational Science and Engineering at Southern University of Science and Technology.

%\newpage

\balance

\bibliographystyle{IEEEtran}
\bibliography{ref}

\end{document}